\documentclass[12pt]{iopart}
\usepackage{bm}

\usepackage{iopams}  
\usepackage{graphicx}
\begin{document}
\def\nn{\nonumber}

\title{disordered topological quantum critical points 
in three-dimensional systems}

\author{Ryuichi Shindou}

\address{Condensed Matter Theory Laboratory, 
RIKEN, 2-1 Hirosawa, Wako, Saitama 351-0198, Japan}
\ead{rshindou@riken.jp} 

\author{Ryota Nakai}
\address{Department of Physics, University of Tokyo, 
7-3-1 Hongo, Bunkyo-ku, Tokyo 113-0033, Japan}
\address{Condensed Matter Theory Laboratory, 
RIKEN, 2-1 Hirosawa, Wako, Saitama 351-0198, Japan}
\ead{rnakai@riken.jp}

\author{Shuichi Murakami}
\address{Department of Physics, 
Tokyo Institute of Technology, 2-12-1 Ookayama, 
Meguro-ku, Tokyo 152-8551, Japan} 
\address{PRESTO, Japan Science and Technology Agency
4-1-8 Honcho, Kawaguchi,
Saitama 332-0012, Japan} 
\ead{murakami@stat.phys.titech.ac.jp}
 
\begin{abstract}
Generic non-magnetic disorder effects onto those  
topological quantum critical points (TQCP),  
which intervene 
the three-dimensional topological insulator and 
an ordinary insulator, are investigated. We first 
show that, in such 3-d TQCP, any backward scattering 
process mediated by the chemical-potential-type impurity 
is always canceled by its 
time-reversal ($\cal T$-reversal) counter-process, 
because of the non-trivial Berry phase supported 
by these two processes in the momentum space. 
However, this cancellation can be generalized  
into {\it only} those backward scattering processes 
which conserve a certain internal degree of freedom, 
i.e. the parity density, while the `absolute' stability 
of the TQCP against {\it any} non-magnetic disorders is 
required by the bulk-edge correspondence.  
Motivated by this, we further derive  
the self-consistent-Born phase diagram in 
the presence of {\it generic} non-magnetic 
disorder potentials and argue the behaviour of 
the quantum conductivity correction in such 
cases. The distinction and similarity between 
the case with only the chemical-potential-type 
disorder and that with the generic 
non-magnetic disorders are finally summarized. 

\end{abstract}

\maketitle

\section{Why non-magnetic disorders in 3-d 
quantum spin Hall systems ?} 
A 3-d $Z_2$ quantum spin Hall 
insulator~\cite{prb-mb,prb-roy,prlb-fkm} is defined to 
be accompanied by the integer number of surface conducting 
channels, each of which is described by the 2+1 
massless Dirac fermion, i.e. helical surface state.  
The stability of each helical 
surface state is protected by the Kramers degeneracy 
at the time-reversal (${\cal T}$-) symmetric (surface) 
crystal momentum. The associated spin is directed within 
the $XY$ plane, and rotates odd-number of times when the  
surface crystal momentum rotates once around this  
${\cal T}$-symmetric $k$-point. 
While an individual helical surface state is 
protected by the ${\cal T}$-symmetry, level-crossings 
between any two different helical surface states are 
generally accidental: they can be lifted 
by a certain ${\cal T}$- 
symmetric perturbation~\cite{prlb-fkm,prl-km,prl-wbz,prb-xm}.  
As a result of this pair annihilation, those 
insulators having even number of helical surface states 
can be adiabatically connected to an ordinary 
band insulator having no surface conducting channels  
at all (`weak topological insulator'). On the other 
hand, those with odd numbers of helical surface states 
are always accompanied by (at least) one gapless 
surface state, even when subjected under this pair 
annihilation (`strong topological insulator').  
In the presence of the ${\cal T}$-symmetry, 
the latter type of insulators cannot be decomposed   
into any two copies of spinless wavefunctions, and 
therefore regarded as a new quantum state of 
matter which goes beyond the quantum Hall paradigm. 

The quantum critical 
point~\cite{njp-m,prb-mk,prb-qtz,prb-srfl} 
which intervenes this 3-d topological insulator and 
a ${\cal T}$-symmetric ordinary band insulator 
is also non-trivial. That is, the stability of 
its critical nature is tightly connected to 
the stability of each helical surface  
state in the topological phase. To be specific,  
consider that a certain ${\cal T}$-symmetric 
model-parameter is changed, 
such that the system transits  
from the topological phase 
to a ${\cal T}$-symmetric 
ordinary phase. During this, 
the helical surface state in the topological phase  
is always protected by the ${\cal T}$-symmetry. 
When the system reaches the quantum critical 
point, however, the bulk-wavefunction becomes extended once. 
Thus, the two helical surface states localized 
at two opposite sample boundaries can 
communicate via this extended bulk-wavefunction, 
only to annihilate with each other. 
Thanks to this pair annihilation 
{\it mediated by the extended bulk}, 
the system can safely remove the stable 
helical surface state and enter the ordinary 
phase having no surface conducting channels, 
while simultaneously keeping the ${\cal T}$-symmetry. 
To put this reversely, 
the bulk-extended character of the 
quantum critical point is required to be stable  
against any ${\cal T}$-symmetric perturbations, 
provided that an individual helical surface state 
in the topological phase is stable against 
the same perturbations. Such a quantum critical 
point can be dubbed as the strong topological 
quantum critical point. 
                
The above picture in the clean limit implies non-trivial 
disorder effects around this topological quantum critical 
point~\cite{prb-sm}. 
To describe this with clarity, let us first define three 
independently-tunable physical 
parameters in the topological phase: (i) the (topological) band 
gap $W_{\rm topo}$, (ii) the band width 
of the conduction/valence band $W_{\rm band}$, and 
(iii) non-magnetic disorder strengths $W_{\rm dis}$. 
Provided that  
$W_{\rm dis}<W_{\rm topo}$, the helical surface state 
in the topological phase is stable against any 
non-magnetic disorders irrespective 
of $W_{\rm band}$~\cite{prl-btbb,prl-rmof,prl-nkr}. 
On increasing $W_{\rm dis}$ such that 
$W_{\rm band}<W_{\rm dis}<W_{\rm topo}$, those bulk-states 
far from the two band-centers, i.e. that of the conduction 
band and of the valence band, become localized 
(see Fig.~1(a))~\cite{prl-aalr}. 
Especially near the zero-energy region, $\mu=0$,   
the system has no extended bulk wavefunctions. 
Thus, two helical surface states localized  
at the two opposite boundaries respectively 
are disconnected from each other. Starting from 
such a localized phase, consider decreasing the 
topological band gap (or increasing the disorder 
strength), such that $W_{\rm band} < W_{\rm topo} < W_{\rm dis}$.
During this decrease, these two disconnected surface 
states should communicate with each other once, so as to  
annihilate with each other, before the system enters 
the ordinary insulator phase. This requires that, at 
$W_{\rm topo} \simeq W_{\rm dis}$, the bulk-wavefunctions  
at $\mu \simeq 0$ must become extended once, so as to mediate 
these two opposite boundaries. 
Moreover, it also requires that such a bulk-extended region 
always intervenes between the topological phase 
and the ordinary insulator phase in the 
three-dimensional parameter space spanned by the chemical 
potential $\mu$, $W_{\rm dis}$ and $W_{\rm topo}$ (see Fig.~1(b,c)). 
Such behaviour of the bulk extended region is often dubbed as  
the levitation and pair-annihilation phenomena, and was 
ubiquitously observed also in other topological 
systems, such as quantum Hall systems~\cite{prb-h,prl-aa,prl-hb} 
and 2-d $Z_2$ quantum spin Hall systems~\cite{prl-oan,prb-ofrm,arx-gwab}.  
\begin{figure}[htp]
\begin{center}
\includegraphics[scale=0.7]{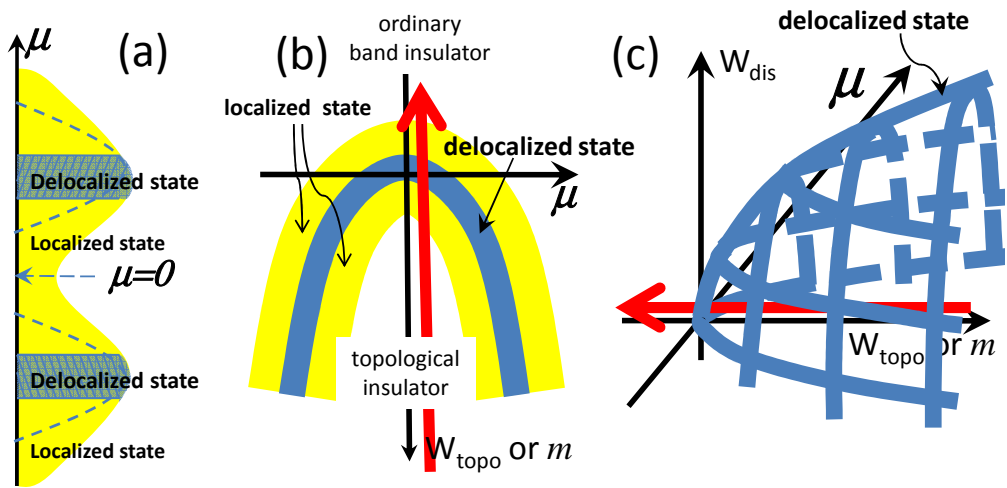}
\caption{ (a) A schematic density of state at 
$W_{\rm dis}<W_{\rm topo}(\equiv m)$. 
(b,c) schematic phase diagrams in the 
$\mu$-$W_{\rm topo}$-$W_{\rm dis}$ plane. The  
blue regions in these three figures stand for the 
delocalized state, which always intervene between the 
topological insulator phase and an ordinary insulator 
phase in any `${\cal T}$-symmetric' phase diagram.  
The figures are derived from Ref.~\cite{prb-sm}. }
\label{spd1}
\end{center}
\end{figure}

Now that the 3-d $Z_2$ topological band insulator 
is a new quantum state of matter having  
no quantum-Hall analogue, the bulk-delocalized phase which  
intervenes this insulating phase (or localized state 
adiabatically connected to this) and the ordinary band  
insulator (or corresponding localized state) is also 
a new type of three-dimensional metallic phase, 
which has no analogue in any other condensed 
matter paradigms. 
Taking into account the chemical-potential type 
disorder (the most representative non-magnetic 
disorder), we have previously studied~\cite{prb-sm} 
the self-consistent Born phase digram 
around this topological metallic phase (or quantum 
critical point) and calculated the 
weak-localization correction to the charge 
conductivity. 
In actual solids~\cite{nat-452-970-08,sci-325-178-09,
np-5-438-09,sci-323-919-09,nat-460-1101-09}, however, 
electronic disorders can generally exist not 
just in the chemical potential, but also  
in the transfer integral and spin-orbit 
interaction potential itself. 
On the one hand, the above argument based on 
the bulk-edge correspondence clearly dictates 
that the critical nature of the topological 
quantum critical point is stable against any of these 
non-magnetic disorders (not just against the chemical 
potential type disorder).  

Motivated by this, we will 
treat in this paper various 
types of non-magnetic disorder potentials 
on an equal footing, so as to argue 
actual/generic properties of 
the topological quantum critical point 
in disordered $Z_2$ quantum spin Hall 
systems. The organization of this paper is as follows. 
In sec.~2, we summarize the 
effective continuum model studied in this paper. 
We next argue in sec.~3 that 
the non-trivial Berry phase (i.e. $e^{i\pi}$) inherent 
in the 3-d topological quantum critical point (TQCP) 
induces the perfect cancellation in the backward 
scattering channels mediated by the chemical-potential 
type disorder. This {\it partially} upholds the above 
argument based on the bulk-edge correspondence. However, 
it also becomes clear that this `Berry phase' argument 
does {\it not necessarily} work in the presence of those 
non-magnetic disorders other than the 
chemical-potential type one. More accurately, 
those backward scattering processes which do not conserve 
a certain internal degree of freedom, i.e. parity density 
degree of freedom, are {\it not generally}   
set off by their ${\cal T}$-reversal 
counter processes. Promoted by this theoretical observation, 
we further derive in sec.~4 
the self-consistent Born phase diagram in the 
presence of {\it generic} ${\cal T}$-symmetric disorder 
potentials. Based on this, we argue in sec.~5 
the weak-localization correction to the charge conductivity in the 
presence of both the chemical-potential type disorder and 
the topological-mass type disorder. 
The distinctions between the case with only the 
chemical potential disorder and that with 
{\it generic} ${\cal T}$-symmetric disorders are 
summarized/clarified in sec.~6.

\section{effective continuum model and ${\cal T}$-symmetric disorders}

An effective continuum model describing the  
topological quantum critical point 
is given by a following $3+1$ Dirac 
fermion~\cite{prlb-fkm,njp-m,prb-mk,prb-qtz}; 
\begin{eqnarray}
{\cal H}_0 \equiv \int d^3 r \psi^{\dagger}(r) 
\bigg\{ \sum_{\mu=1}^3 \hat{\gamma}_{\mu} (-i\partial_{\mu}) 
+ m \hat{\gamma}_5 \bigg\} \psi(r),  \label{ham} 
\end{eqnarray}  
where the two massive phases, $m>0$ and $m<0$, represent the 
quantum spin Hall insulator and the ${\cal T}$-symmetric 
conventional (band) insulator respectively: `$m$' corresponds to 
the topological band gap $W_{\rm topo}$. The $4$ by $4$ 
$\gamma$-matrices consist of the Pauli spin matrix part ${\bm s}$ and 
the other Pauli matrix ${\bm \sigma}$ representing the sublattice 
(or orbital) degree of freedom. There exist at most five 
such $\gamma$-matrices which anti-commute 
with one another. Here they are dubbed as 
$\gamma_{1}$, $\gamma_2$, 
$\gamma_3$, $\gamma_4$ and $\gamma_5$. 
The product of these five always reduces 
to (the minus of) the unit matrix;  
$\hat{\gamma}_1 \hat{\gamma}_2 \hat{\gamma}_3 
\hat{\gamma}_4 \hat{\gamma}_5 = -1$. 
Thus, even numbers of $\gamma$-matrices out of these five 
should be time-reversal (${\cal T}$)-odd and spatial 
inversion (${\cal I}$)-odd. In eq.~(\ref{ham}), three of 
them are already linearly coupled with the momentum. 
Thus, four of them should be both ${\cal T}$-odd and ${\cal I}$-odd. 
Without loss of generality, eq.~(\ref{ham}) took 
$\gamma_{1,2,3,4}$ to be ${\cal T}$-odd and ${\cal I}$-odd, 
so that $\gamma_{5}$ is ${\cal T}$-even and ${\cal I}$-even. 
Under this convention, $\gamma_{15} \equiv -i\gamma_1\gamma_5$, 
$\gamma_{25}$, $\gamma_{35}$ and 
$\gamma_{45}$ are all ${\cal T}$-even and 
${\cal I}$-odd, while $\gamma_{12}$, $\gamma_{13}$, $\gamma_{14}$, 
$\gamma_{23}$, $\gamma_{34}$ and $\gamma_{42}$ are all 
${\cal T}$-odd and ${\cal I}$-even.  
Thus, arbitrary `on-site' type  non-magnetic disorder 
potentials are generally given by, 
\begin{eqnarray}
{\cal H}_{\rm imp} \equiv \int d^3r 
\psi^{\dagger}(r) \bigg\{ v_0(r) \hat{\gamma}_0 
+ v_5 (r) \hat{\gamma}_5 + 
\sum_{j=1}^4 v_{j5}(r) \hat{\gamma}_{j5} \bigg\} \psi(r). \label{imp} 
\end{eqnarray} 
with the real-valued functions $v_{j}$ $(j=0,5,15,\cdots,45)$.  

In Ref.~\cite{prb-sm}, we have further assumed 
that the chemical-potential type disorders is the most dominant 
and have set the other five to be zero: 
$v_{5}=v_{15}=\cdots = v_{45}=0$.  
Based on these simplifications, we have carried out the 
self-consistent Born analysis and weak-localization 
calculations, and obtained a simple microscopic picture, 
which can indeed explain how the bulk-extended region 
emerges when the topological band gap `$m$' changes its 
sign~\cite{prb-sm}. In the next section, we will 
describe an alternative argument, which more directly  
dictates that the topological metallic phase 
(or quantum critical point) is free from any localization 
effect induced by generic non-magnetic disorders. 

\section{`Berry phase' effect in 3-d topological metallic phases}
Our argument here is a straightforward 3-dimensional generalization of 
the `Berry phase' argument invented by Ando 
{\it et al.}~\cite{jpsj-ans} in the context of the 
carbon-nanotube (or 2-d graphene) subjected 
under {\it long}-range impurity potentials. 
Specifically, we will prove that an individual backward 
scattering process is precisely canceled by its time-reversal 
counterpart, which leads to a complete absence of the backward 
scattering in the 3-d topological quantum critical point. 

Consider first the most simplest case, where the $T$-matrix is 
composed only by the chemical-potential-type impurity  $v_0(r)$,  
\begin{eqnarray}
\hspace{-1cm} 
\hat{T} \equiv v_0
 + v_0
\frac{1}{\epsilon-\hat{\gamma}_\mu (-i\partial_{\mu})} 
v_0
+ v_0
\frac{1}{\epsilon-\hat{\gamma}_\mu (-i\partial_{\mu})} 
v_0
\frac{1}{\epsilon-\hat{\gamma}_\mu (-i\partial_{\mu})} 
v_0
+ \cdots . \nn 
\end{eqnarray} 
In the momentum representation, the $(p+1)$-th order backward scattering term reads, 
\begin{eqnarray}
&& \hspace{-2.5cm}
\langle -k,\epsilon_{-k},\sigma|\hat{T}^{(p+1)}|k,\epsilon_{k},\sigma'\rangle 
= \sum_{\tau_1,k_1,\sigma_1}\cdots \sum_{\tau_p,k_p,\sigma_p} 
 \frac{v_0(-k-k_{p})\cdots v_{0}(k_2-k_1) v_0(k_1-k)}
{(\epsilon-\tau_{p}\epsilon_{k_{p}})\cdots
(\epsilon-\tau_{2}\epsilon_{k_{2}})
(\epsilon-\tau_{1}\epsilon_{k_{1}})}\times  \nn \\ 
&& \hspace{-1.8cm} 
\langle - k,\epsilon_{-k},\sigma|k_{p},\tau_{p}\epsilon_{k_p}, 
\sigma_{p} \rangle \times \cdots \times 
\langle k_{2},\tau_{2}\epsilon_{k_2}, 
\sigma_{2} | k_{1},\tau_{1}\epsilon_{k_1},\sigma_1\rangle 
\langle k_1,\tau_1\epsilon_{k_1},\sigma_1|
k,\epsilon_k,\sigma'\rangle,   \label{mat}
\end{eqnarray} 
where $\tau_j=\pm$ ($j=1,\cdots,p$) in 
combination with $\epsilon_{k_j}$ 
specifies the eigen-energy of the hamiltonian at the 
momentum $k_j$, i.e. $\tau_j \epsilon_{k_j} = \pm |k_j|$.  
Such an eigen-energy is always accompanied by 
doubly degenerate eigen-states, which is specified 
by the spin index, $\sigma_j=\uparrow,\downarrow$. 
Namely, each eigen-state at the momentum $k_j$ is 
uniquely identified by these two indices, 
$|k_j,\tau_j\epsilon_{k_j},\sigma_j\rangle$. Without  
loss of generality, we can take the initial ket-state 
and final bra-state to have an positive eigen-energy, 
i.e. $|k,\epsilon_k,\sigma\rangle$ and 
$\langle -k,\epsilon_{-k},\sigma|$. 
We can also fix the gauge of the eigen-wavefunctions, e.g. 
\begin{eqnarray}
|k,\tau \epsilon_{k_j},\sigma\rangle \equiv  
e^{i\frac{\theta}{2}\big(\sin\phi\hat{\gamma}_{23} - 
\cos\phi \hat{\gamma}_{31}\big)} |\tau,\sigma\rangle_3 \equiv 
\hat{U}_{k} |\tau,\sigma\rangle_3, \nn  
\end{eqnarray} 
with $k=|k|(\sin\theta\cos\phi,\sin\theta\sin\phi,\cos\theta)$. 
$|\tau,\sigma\rangle_3$ is the two-fold degenerate eigenstates of 
$\hat{\gamma}_3$ which belong to its eigenvalue $\tau$, i.e. 
$\hat{\gamma}_3 |\tau,\sigma\rangle_3 \equiv 
\tau |\tau,\sigma\rangle_3$.    

Based on this set-up, we can explicitly argue that the 
following two backward scattering processes, which 
are related by time-reversal operation, set off each 
other in eq.~(\ref{mat}),  
\begin{eqnarray}
&&\hspace{-0.3cm} 
|k,+,\sigma \rangle \rightarrow |k_1,\tau_1,\cdots\rangle  
\rightarrow \cdots \rightarrow |k_p,\tau_{p},\cdots \rangle 
\rightarrow |-k,+,\sigma'\rangle, \nn \\ 
&&\hspace{-0.3cm} 
|k,+,\sigma \rangle \rightarrow |-k_p,\tau_{p},\cdots\rangle  
\rightarrow \cdots  \rightarrow |-k_1,\tau_1,\cdots\rangle   
\rightarrow |-k,+,\sigma'\rangle. \nn  
\end{eqnarray}
where the summation over the spin indices associated with the 
intermediate states are assumed. To be more explicit, we can 
show that the following two have an opposite sign with each other,
\begin{eqnarray}
&& \hspace{-0.2cm}
S^{(p+1)}_{\sigma\sigma^{\prime}}  
= \sum_{\{\sigma_j\}}   
\langle - k,\epsilon_{k},\sigma|k_{p},\tau_{p}\epsilon_{k_p}, 
\sigma_{p} \rangle \cdots  
\langle k_1,\tau_1\epsilon_{k_1},\sigma_1|
k,\epsilon_k,\sigma' \rangle, \nn \\ 
&&\hspace{-0.2cm}
S^{(p+1)\prime}_{\sigma\sigma^{\prime}} = \sum_{\{\sigma_j\}} 
\langle - k,\epsilon_{k},\sigma|-k_{1},\tau_{1}\epsilon_{k_1}, 
\sigma_{1} \rangle \cdots  \langle -k_{p},\tau_{p}\epsilon_{k_p}, 
\sigma_{p} | k,\epsilon_{k},\sigma'\rangle. \nn 
\end{eqnarray}
for any $\sigma$ and $\sigma'$. To see this, 
take the sum over all the intermediate spin indices 
first. 
This leads to 
\begin{eqnarray}
&& \hspace{-2.cm} 
S^{(p+1)}_{\sigma\sigma^{\prime}} =    
\ _3 \langle +,\sigma| \hat{U}^{\dagger}_{-k} 
\cdot (\hat{\gamma}_0 - n_{p,\mu} \hat{\gamma}_{\mu}) 
\cdots (\hat{\gamma}_0 -  n_{2,\nu} \hat{\gamma}_{\nu}) 
\cdot 
(\hat{\gamma}_0 - n_{1,\lambda} \hat{\gamma}_{\lambda}) 
\cdot \hat{U}_k |+,\sigma'\rangle_3. \nn \\    
&&\hspace{-2.cm}
S^{(p+1)\prime}_{\sigma\sigma^{\prime}} =\ _3 \langle +,\sigma| 
\hat{U}^{\dagger}_{-k} 
\cdot (\hat{\gamma}_0 + n_{1,\mu} \hat{\gamma}_{\mu}) 
\cdot (\hat{\gamma}_0 + n_{2,\nu} \hat{\gamma}_{\nu}) 
\cdots 
(\hat{\gamma}_0 + n_{p,\lambda} \hat{\gamma}_\lambda) \cdot 
\hat{U}_k |+,\sigma'\rangle_3. 
\nn 
\end{eqnarray} 
where $n_j$ is a product between the normalized vector parallel
to $k_j$ and $\tau_j$, $n_j \equiv \tau_j k_j/|k_j|$.  
The sequential product of the projection operators can be 
further expanded in this unit vector, i.e.  
\begin{eqnarray}
&& \hspace{-2.2cm} 
(1+n_p\cdot \gamma)\cdots (1+n_2\cdot \gamma)\cdot 
(1+ n_1\cdot \gamma) \nn\\
&&\hspace{-0.7cm} 
= 1 + \sum n_j \cdot \gamma 
+ \sum_{i>j} (n_i \cdot \gamma) (n_j \cdot \gamma) 
+ \cdots + (n_p \cdot \gamma) \cdots (n_{2}\cdot \gamma)
(n_1\cdot \gamma). \label{exp} 
\end{eqnarray} 
An even-order term in this expansion is 
always a linear combination of $\gamma_0$, $\gamma_{23}$, 
$\gamma_{31}$ and $\gamma_{12}$,  
\begin{eqnarray}
\hspace{0.7cm}
(n_{2l}\cdot \gamma)\cdots (n_{1}\cdot \gamma) 
= A_{l} \gamma_0 + \epsilon_{\mu\nu\lambda}B_{l,\mu} 
\gamma_{\nu\lambda}. \nn
\end{eqnarray}
$A_{l}$ is a scalar quantity, which is composed 
only by inner products among $2l$ unit vectors. 
On the other hand, $B_l$ clearly behaves as a vector, being 
coupled to $\gamma_{23}$, $\gamma_{31}$ and $\gamma_{12}$. 
It is expressed as (a linear combination of) the product 
between $l-1$ inner products and one outer product,  
\begin{eqnarray} 
B_{l,\mu} \equiv  \sum_{l,m,\cdots,n,o\ne i,j} 
b^{l}_{ij,lm\cdots no} (n_i \times n_j)_{\mu} (n_l\cdot n_m) \cdots 
(n_{n}\cdot n_{o}),  \nn 
\end{eqnarray}  
with some coefficients $b^{\cdots}_{\cdots}$ free from $\{n_1,\cdots,n_{2l}\}$. 
Multiply this even-order term by one additional $n \cdot \gamma$, 
one can obtain any odd-order term in eq.~(\ref{exp}) as a 
linear combination of $\gamma_1$, $\gamma_2$,  
$\gamma_3$ and $\gamma_{45}$; 
\begin{eqnarray} 
\hspace{0.7cm}
(n_{2l+1}\cdot \gamma)\cdots (n_{1}\cdot \gamma) 
= C_{l,\mu} \gamma_{\mu} + D_{l}\gamma_{45}. \nn 
\end{eqnarray} 
$C_{l}$ is a vector composed of $l$ inner products, 
while $D_l$ is a scalar containing one scalar triple product, 
\begin{eqnarray} 
&&\hspace{-0.2cm} 
C_{l,\mu} = \sum_{n,o,\cdots,p,q\ne m} c^{l}_{m,no\cdots pq} 
n_{m,\mu} (n_n\cdot n_o)\cdots (n_p\cdot n_q), \nn \\  
&&\hspace{-0.2cm}  
D_l = \sum_{n,0\cdots,p,q \ne i,j,m} d^l_{ijm,no\cdots pq}
(n_i \times n_j)\cdot n_m (n_n\cdot n_o)\cdots (n_p\cdot n_q). \nn  
\end{eqnarray} 
Thus, under the time-reversal operation, i.e. 
$(n_p,n_{p-1},\cdots,n_1) \rightarrow (-n_1,-n_2,\cdots,-n_p)$, 
$A_l$ and $D_l$ do not change the sign, while $B_l$ and $C_l$ 
change their signs. These observations lead to a perfect 
cancellation between $S^{(p+1)}_{\sigma\sigma'}$ and 
$S^{(p+1)\prime}_{\sigma\sigma'}$, 
\begin{eqnarray}
\hspace{-1.6cm} 
S^{(p+1)}_{\sigma\sigma^{\prime}} + 
S^{(p+1)\prime}_{\sigma\sigma^{\prime}} 
&=& 2{\sum}\ _3 \langle +,\sigma|\hat{U}^{\dagger}_{-k} 
\cdot \big\{A \hat{\gamma}_0 + D \hat{\gamma}_{45} \big\}\cdot  \hat{U}_k 
|+,\sigma^{\prime}\rangle_3, \nn \\ 
&=& 2i{\sum}\  _3 \langle +,\sigma| (-\sin\phi \hat{\gamma}_{23} 
+ \cos\phi \hat{\gamma}_{31})\cdot ( A \hat{\gamma}_0 + D \hat{\gamma}_{45}) 
|+,\sigma^{\prime}\rangle_3 \nn \\ 
&=& \sum_{\sigma^{\prime\prime}} 
(\cdots) \cdot \  _3 \langle +,\sigma|-,\sigma^{\prime\prime} 
\rangle_3 = 0, \nn  
\end{eqnarray}  
for arbitrary $\sigma$ and $\sigma'$. 
This dictates that the backward-scattering process mediated 
by the chemical-potential-type disorder vanishes completely 
at the 3-d TQCP. 

More generally, we can prove that any 
backward scattering process {\it which conserves the eigenvalue 
of $\hat{\gamma}_{45}$} is always set off by its corresponding 
${\cal T}$-reversal counterpart. Such a process is always 
mediated by {\it even} number of either 
$v_{j5}(r)$ ($j=1,2,3$) or $v_5(r)$, since  
both $\hat{\gamma}_{j5}$ ($j=1,2,3$) and 
$\hat{\gamma}_{5}$ anticommute with 
$\hat{\gamma}_{45}$. To uphold the complete absence of the backward scattering 
in this case, we have only to 
show that the following two matrix elements  
have an opposite sign with each other, 
\begin{eqnarray}
&&\hspace{-2.3cm}
S^{(p+1)}_{\sigma\sigma^{\prime}} 
= {_3} \langle +,\sigma| \hat{U}^{\dagger}_{-k} 
\hat{\gamma}_{a_p}  
(\hat{\gamma}_0 - n_{p,\mu}\hat{\gamma}_{\mu})    
\cdots \hat{\gamma}_{a_2}(\hat{\gamma}_0 - 
n_{2,\nu}\hat{\gamma}_{\nu}) \hat{\gamma}_{a_1} 
(\hat{\gamma}_0 - n_{1,\lambda}\hat{\gamma}_{\lambda}) 
\hat{\gamma}_{a_0} \hat{U}_k |+,\sigma^{\prime} \rangle_3 \nn \\   
&&\hspace{-2.3cm}
S^{(p+1)\prime}_{\sigma\sigma^{\prime}} 
= {_3} \langle +,\sigma| \hat{U}^{\dagger}_{-k} 
\hat{\gamma}_{a_0}  
(\hat{\gamma}_0 + n_{1,\mu}\hat{\gamma}_{\mu})  
\hat{\gamma}_{a_1}    
(\hat{\gamma}_0 +  
n_{2,\nu}\hat{\gamma}_{\nu}) 
\hat{\gamma}_{a_2} \cdots 
(\hat{\gamma}_0 + n_{p,\lambda}\hat{\gamma}_{\lambda}) 
\hat{\gamma}_{a_p} 
\hat{U}_k |+,\sigma^{\prime} \rangle_3, \nn   
\end{eqnarray}  
for arbitrary $\sigma$ and $\sigma^{\prime}$. 
The indices $a_0$, $a_1,a_2,\cdots $ and $a_p$ can be either   
$0$, $15$, $25$, $35$ , $45$ or $5$, under the condition 
that the total number of those $\hat{\gamma}_{15}$, $\hat{\gamma}_{25}$, 
$\hat{\gamma}_{35}$ and $\hat{\gamma}_5$ which mediate 
the initial state and the final state is always {\it even}. To 
see the relative sign between these two, 
let us first transport all the intervening 
$\gamma_{a_j}$s in the leftward/rightward until they meet with 
the bra/ket states in 
$S^{(p+1)}_{\sigma\sigma^{\prime}}$
/$S^{(p+1)\prime}_{\sigma\sigma^{\prime}}$. 
This transport brings about an appropriate 
redefinition of the normalized vectors, 
$n_{j} \rightarrow \overline{n}_j$, 
\begin{eqnarray}
&& \hspace{-1.5cm}
S^{(p+1)}_{\sigma\sigma^{\prime}} = {_3} 
\langle +,\sigma| \hat{U}^{\dagger}_{-k}
\hat{\gamma}_{a_p}\cdots 
\hat{\gamma}_{a_1}\hat{\gamma}_{a_0} \cdot 
(\hat{\gamma}_0-\overline{n}_{p,\mu}\hat{\gamma}_{\mu})
\cdots
(\hat{\gamma}_0-\overline{n}_{1,\lambda}\hat{\gamma}_{\lambda})
\hat{U}^{\dagger}_k|+,
\sigma^{\prime}\rangle_3 \nn \\ 
&&\hspace{-1.5cm} 
S^{(p+1)\prime}_{\sigma\sigma^{\prime}} = {_3} 
\langle +,\sigma| \hat{U}^{\dagger}_{-k} 
(\hat{\gamma}_0+\overline{n}_{1,\mu}\hat{\gamma}_{\mu})
\cdots(\hat{\gamma}_0+\overline{n}_{p,\lambda}\hat{\gamma}_{\lambda})
\cdot 
\hat{\gamma}_{a_0}\hat{\gamma}_{a_1}\cdots 
\hat{\gamma}_{a_p}
\hat{U}_k |+,
\sigma^{\prime}\rangle_3. \nn 
\end{eqnarray}
$S^{(p+1)}$ and $S^{(p+1)\prime}$ is still  
connected by the ${\cal T}$-reversal operation, 
$(\overline{n}_p,\overline{n}_{p-1},\cdots,\overline{n}_1) 
\leftrightarrow (-\overline{n}_1,\cdots,-\overline{n}_{p-1},
-\overline{n}_p)$.  Thus, the preceding 
expansion can be used again, 
\begin{eqnarray}
&&\hspace{-2cm} 
(\hat{\gamma}_0-\overline{n}_{p,\mu}\hat{\gamma}_{\mu})
\cdots(\hat{\gamma}_0 -\overline{n}_{2,\nu}\hat{\gamma}_{\nu})
(\hat{\gamma}_0-\overline{n}_{1,\lambda}\hat{\gamma}_{\lambda})  
= \overline{A} \hat{\gamma}_0 + 
\overline{B}_{\mu}\epsilon_{\mu\nu\lambda}
\hat{\gamma}_{\nu\lambda} + 
\overline{C}_{\mu}\hat{\gamma}_{\mu} + \overline{D} 
\hat{\gamma}_{45}, \nn \\ 
&&\hspace{-2cm} 
(\hat{\gamma}_0+\overline{n}_{1,\mu}\hat{\gamma}_{\mu})
(\hat{\gamma}_0 + \overline{n}_{2,\nu}\hat{\gamma}_{\nu})
\cdots 
(\hat{\gamma}_0 + \overline{n}_{p,\lambda}\hat{\gamma}_{\lambda})  
= \overline{A} \hat{\gamma}_0 -  
\overline{B}_{\mu}\epsilon_{\mu\nu\lambda}
\hat{\gamma}_{\nu\lambda} -  
\overline{C}_{\mu}\hat{\gamma}_{\mu} + \overline{D} 
\hat{\gamma}_{45}. \nn 
\end{eqnarray}

\begin{figure}[h]
\begin{center}
\includegraphics[scale=0.65]{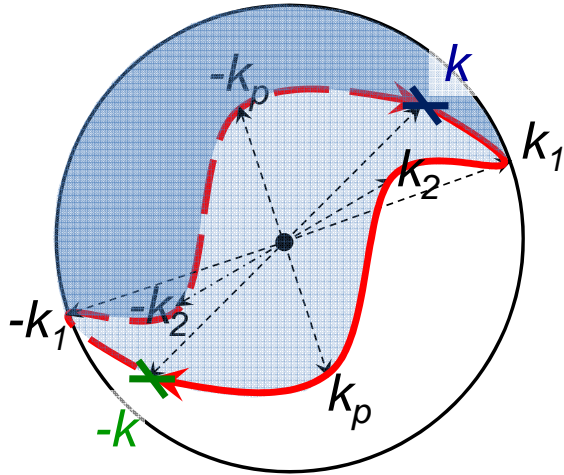}
\caption{Red curves denote 
two backward scattering processes on a sphere 
in the momentum space ($|k_1|=\cdots=|k_p|=|k|$), 
which are mutually time-reversal to 
each other. The blue-shaded region, whose boundary 
is shaped by this two path, occupies {\it one half} of 
the whole surface of a sphere. Thus, an electron 
acquires the Berry phase $\pi$ (instead of $2\pi$), when 
it travels once around this boundary.}
\label{pd0}
\end{center}
\end{figure}

The condition imposed on $a_j$ dictates 
that $\hat{\gamma}_{a_p}\cdots \hat{\gamma}_{a_1}
\hat{\gamma}_{a_0}$ and $\hat{\gamma}_{a_0}\hat{\gamma}_{a_1}
\cdots \hat{\gamma}_{a_p}$ always reduce to  
either (i) $\hat{\gamma}_{j}$ and $-\hat{\gamma}_j$ 
($j=1,2,3,12,23,31$) respectively or (ii) 
$\hat{\gamma}_{m}$ and $\hat{\gamma}_{m}$ ($m = 0,45$) 
respectively. 
Consider the former case first, e.g. 
\begin{eqnarray}
&&\hspace{-1cm} 
S^{(p+1)}_{\sigma\sigma^{\prime}} 
=  {_3} 
\langle +,\sigma| \hat{U}^{\dagger}_{-k}
\cdot \hat{\gamma}_{1}\cdot 
(\overline{A} \hat{\gamma}_0 +  
\overline{B} \hat{\gamma}_{\nu\lambda} +  
\overline{C}\hat{\gamma}_{\mu} + \overline{D} 
\hat{\gamma}_{45})\cdot 
\hat{U}_k|+,
\sigma^{\prime}\rangle_3 \nn \\ 
&&\hspace{0.1cm} = {_3} 
\langle +,\sigma| \hat{U}^{\dagger}_{-k}
\hat{U}_k 
\cdot m_{\rho}\hat{\gamma}_{\rho}\cdot 
(\overline{A} \hat{\gamma}_0 +  
\overline{B}^{\prime} \hat{\gamma}_{\nu\lambda} +  
\overline{C}^{\prime} \hat{\gamma}_{\mu} 
+ \overline{D} 
\hat{\gamma}_{45})|+,
\sigma^{\prime}\rangle_3, \nn \\ 
&&\hspace{-1cm} 
S^{(p+1)\prime}_{\sigma\sigma^{\prime}} 
= {_3} 
\langle +,\sigma| \hat{U}^{\dagger}_{-k}\cdot 
(-\overline{A} \hat{\gamma}_0 +    
\overline{B} \hat{\gamma}_{\nu\lambda} +    
\overline{C}\hat{\gamma}_{\mu} - \overline{D} 
\hat{\gamma}_{45})\cdot 
\hat{\gamma}_{1}\cdot 
\hat{U}_k|+,
\sigma^{\prime}\rangle_3 \nn \\
&&\hspace{0.1cm} 
= {_3} 
\langle +,\sigma| \hat{U}^{\dagger}_{-k}
\hat{U}_k \cdot 
(-\overline{A} \hat{\gamma}_0 +    
\overline{B}^{\prime} \hat{\gamma}_{\nu\lambda} +    
\overline{C}^{\prime}\hat{\gamma}_{\mu} 
- \overline{D} \hat{\gamma}_{45})\cdot 
m_{\rho}\hat{\gamma}_{\rho} |+,
\sigma^{\prime}\rangle_3. \nn
\end{eqnarray} 
with $\hat{U}^{\dagger}_{-k}\hat{U}_{k} = -\sin\phi \hat{\gamma}_{23} 
+ \cos\phi \hat{\gamma}_{31}$ and 
$m_{\rho}\hat{\gamma}_{\rho} \equiv  
\hat{U}^{\dagger}_k \hat{\gamma}_1 \hat{U}_k$. 
When $\hat{\gamma}_{a_p}\cdots 
\hat{\gamma}_{a_2}\hat{\gamma}_{a_1} = \hat{\gamma}_{12}$ 
(or $\hat{\gamma}_{23}$, $\hat{\gamma}_{31}$),  
$m_{\rho}\hat{\gamma}_{\rho}$ should be replaced by 
$m_{\rho}\epsilon_{\rho\phi\psi}\hat{\gamma}_{\phi\psi}$. 
In either cases, $S^{(p+1)}_{\sigma\sigma^{\prime}}$ 
and $S^{(p+1)\prime}_{\sigma\sigma^{\prime}}$  
set off each other completely e.g.   
\begin{eqnarray} 
&& \hspace{-2cm}
\hat{S}^{(p+1)}_{\sigma\sigma^{\prime}} 
+ \hat{S}^{(p+1)\prime}_{\sigma\sigma^{\prime}} 
= {_3} 
\langle +,\sigma| (-s_{\phi} \gamma_{23} + c_{\phi} \gamma_{31}) 
\cdot 
\{     
\overline{B}^{\prime} \hat{\gamma}_{\nu\lambda} 
+  \overline{C}^{\prime}\hat{\gamma}_{\mu},  
m_{\rho}\hat{\gamma}_{\rho}\} |+,
\sigma^{\prime}\rangle_3 \nn  \\ 
&&\hspace{0.74cm} 
= 2\times {_3} 
\langle +,\sigma| (-s_{\phi} \gamma_{23} + c_{\phi} \gamma_{31}) 
\cdot (\overline{B}^{\prime}_{\mu}m_{\mu} \hat{\gamma}_{45} 
+ \overline{C}^{\prime}_{\nu}m_{\nu} \hat{\gamma}_0) 
|+,\sigma^{\prime}\rangle_3 \nn \\
&& \hspace{0.74cm} = (\cdots)\times  {_3} \langle - ,\sigma^{\prime\prime}| 
+, \sigma^{\prime} \rangle_3 = 0. \nn   
\end{eqnarray}   
A similar cancellation also holds true for the case with (ii)
$\hat{\gamma}_{a_p}\cdots 
\hat{\gamma}_{a_2}\hat{\gamma}_{a_1} 
= \hat{\gamma}_{a_1} \hat{\gamma}_{a_2} \cdots 
\hat{\gamma}_{a_p} = \hat{\gamma}_{45}$. 
These observations thus conclude that any backward scattering process which 
conserves the eigenvalue of $\gamma_{45}$ is always canceled  
by its time-reversal counter-process. 
Like in the 2-d graphene 
case~\cite{jpsj-ans}, this cancellation is 
actually a direct consequence of 
the non-trivial  Berry phase `$e^{i\pi}$' accumulated 
along a closed loop composed by two time-reversal 
counterparts in the momentum space 
(see Fig.~\ref{pd0}).    

The above 'Berry phase' argument, however, does not hold 
true for the case with those backward scattering processes  
which change the eigenvalues  
of $\hat{\gamma}_{45}$. To see this, consider, for example, the  
backward scattering processes mediated by {\it odd} number 
of $v_5(r)$ and compare following two 
${\cal T}$-reversal paired processes,  
\begin{eqnarray}
&& \hspace{-2.1cm} 
S^{(p+1)}_{\sigma\sigma^{\prime}} =    
\ _3 \langle +,\sigma| \hat{U}^{\dagger}_{-k} \cdot 
\hat{\gamma}_5  
\cdot (\hat{\gamma}_0 - n_{p,\mu} \hat{\gamma}_{\mu}) 
\cdots (\hat{\gamma}_0 -  n_{2,\nu} \hat{\gamma}_{\nu}) 
\cdot 
(\hat{\gamma}_0 - n_{1,\lambda} \hat{\gamma}_{\lambda}) 
\cdot \hat{U}_k |+,\sigma'\rangle_3, \nn \\  
&&\hspace{-2.1cm}
S^{(p+1)\prime}_{\sigma\sigma^{\prime}} =\ _3 \langle +,\sigma| 
\hat{U}^{\dagger}_{-k} 
\cdot (\hat{\gamma}_0 + n_{1,\mu} \hat{\gamma}_{\mu}) 
\cdot (\hat{\gamma}_0 + n_{2,\nu} \hat{\gamma}_{\nu}) 
\cdots 
(\hat{\gamma}_0 + n_{p,\lambda} \hat{\gamma}_\lambda) \cdot 
\hat{\gamma}_5 \cdot 
\hat{U}_k |+,\sigma'\rangle_3.  
\nn 
\end{eqnarray}
We have already made any two `neighboring' $\hat{\gamma}_5$ to 
annihilate each other, which leads an appropriate redefinition 
of the normalized unit vectors.  
The extra $\hat{\gamma}_5$ was then transported 
leftward/rightward in $S^{(p+1)}$/$S^{(p+1)\prime}$, 
until it meets with the bra/ket-state.  
Exploiting the expansion described above, we can readily see    
that these two do not cancel exactly in general, 
\begin{eqnarray}
&&\hspace{-2.2cm} 
S^{(p+1)}_{\sigma\sigma'} + 
S^{(p+1)\prime}_{\sigma\sigma'} 
= 2i\sum \ _3\langle +, \sigma| (-\sin\phi \hat{\gamma}_{23} 
+ \cos\phi \hat{\gamma}_{31})\cdot (A + \overline{C}_3 \hat{\gamma}_{3}) 
\cdot \hat{\gamma}_5 | +, \sigma'\rangle_3 \ne 0. \nn  
\end{eqnarray}   

This situation is reminiscent 
of the 2-d graphene (or carbon-nanotube) 
subjected under {\it short}-range impurity potentials, where 
the {\it intervalley} scattering\footnote{which corresponds to 
the {\it inter-'eigenvalue'} scattering in the current context.} suppresses 
the cancellation due to 
the Berry phase and consequently induces the crossover from 
the symplectic class to the orthogonal class~\cite{jpsj-ans,jpsj-as}.   
The situation in the 3-d TQCP is, however, quite different 
from this 2-d analogue. First of all, the preceding argument based 
on the bulk-edge correspondence requires the extended character 
of the 3-d TQCP to be stable against {\it any} non-magnetic disorders. 
Moreover, it always belongs to 
the symplectic class irrespectively of the type of non-magnetic 
disorders, which was not the case in the 
2-d graphene~\cite{jpsj-as}. Therefore, 
not only from the actual material's 
viewpoint~\cite{nat-452-970-08,sci-325-178-09,
np-5-438-09,sci-323-919-09,nat-460-1101-09}, 
but also from the theoretical standpoint, it is quite 
intriguing and important to investigate the effect 
of {\it general} ${\cal T}$-symmetric 
disorder potentials (including the topological-mass 
type one) in the 3-d topological metallic phase.  
In the next section, we will first study how a single-body 
Green function is renormalized by these {\it general} 
non-magnetic disorders. Based on this analysis, we 
will argue the behaviour of the quantum conductivity correction 
in the presence of both the chemical-potential type 
disorder $v_0$ and topological-mass type disorder $v_5$. 
       
\section{self-consistent Born phase diagram -- general case --} 

Consider ${\cal H}_{0}+{\cal H}_{\rm imp}$ with randomly 
distributed real-valued $v_{j}(r)$ ($j=0,5,\cdots,45$). 
Because of the ${\cal T}$-reversal 
symmetry, each eigenstate of this, say $|\psi_n (r) \rangle$, 
has its own Kramers-paired  state $| \overline{\psi}_n (r) \rangle$ 
: $\langle \overline{\psi}_n (r)| \equiv -i\hat{s}_y |\psi_n(r) \rangle$. 
Thus, the single-body Green functions obey 
the following relation; 
\begin{eqnarray}
\hspace{-1cm}\hat{G}^{R(A)}(r,r';\mu) 
&\equiv& \sum_{n} \frac{| \psi_n (r) \rangle \langle \psi(r') |}
{\mu-\epsilon_n \pm i\delta} 
= \hat{s}_y  
\big\{\hat{G}^{R(A)}(r',r;\mu)\big\}^t \hat{s}_y, \label{tsymmetry} 
\end{eqnarray}
where `$+/-$' sign is for the retarded/advanced Green function 
$\hat{G}^{R/A}$. 

The quenched disorder average is taken at the Gaussian level with 
the short-ranged correlation; 
\begin{eqnarray}  
\overline{\cdots} \equiv \frac{1}{\cal N}\int {\cal D}[v] 
\cdots \exp\bigg[-\sum_{j,l} \int d^3 r \Delta_{j,l} v_j (r) v_l (r)\bigg]. 
\label{que}   
\end{eqnarray} 
Green functions thus averaged acquire the   
translational symmetry,  
$\hat{G}^{R(A)}(r,r';\mu)=\hat{G}^{R(A)}(r-r';\mu)$.  
When Fourier-transformed, they can be expanded in terms of 
the 16 $\gamma$-matrices  
with some complex-valued coefficients, 
\begin{eqnarray}
\hspace{-2.2cm} 
\hat{G}^{R}(k;\mu) = \sum_{n=1}^{4} \hat{\gamma}_n   
\overline{\sf F}_{n}(k;\mu) + \sum_{j=12,13,\cdots}^{42} \hat{\gamma}_{j} 
\overline{\sf F}_{j}(k;\mu) + \sum_{m=0,5} \hat{\gamma}_m 
\overline{\sf F}_{m}(k;\mu) + \sum_{l=15}^{45} \hat{\gamma}_{l} 
\overline{\sf F}_{l}(k;\mu),  \nn
\end{eqnarray} 
and $\hat{G}^{A}(k;\mu)=\{\hat{G}^{R}(k;\mu)\}^{\dagger}$.
The ${\cal T}$-symmetry requires 
$\overline{\sf F}_{1,2,3,4}(k)$ and 
$\overline{\sf F}_{12,13,14,23,34,42}(k)$ to be  
odd functions of $k$, while 
$\overline{\sf F}_{0,5}(k)$ and 
$\overline{\sf F}_{15,25,35,45}(k)$ 
to be even in $k$. Thus, only the latter six 
participate in the self-consistent 
Born equation,   
\begin{eqnarray}
&&\hspace{-0.85cm} 
\hat{\Sigma}^{R}(\mu) \equiv 
\hat{G}^{R,-1}_0 (k;\mu) - \hat{G}^{R,-1}(k;\mu) 
= \sum_{j,l} \Delta_{j,l} \int dk' \hat{\gamma}_j \cdot  
\hat{G}^{R}(k';\mu) \cdot \hat{\gamma}_l, \nn \\ 
&&\hspace{0.17cm} 
= \sum_{j,l} \Delta_{j,l} \int dk' \hat{\gamma}_j \cdot 
\Big\{\hat{\gamma}_0 \overline{\sf F}_0(k') + 
\hat{\gamma}_5 \overline{\sf F}_5(k') + 
\sum_{n=1}^{4} \hat{\gamma}_{n5} \overline{\sf F}_{n5}(k')
\Big\} \cdot \hat{\gamma}_l. \label{scb1}  
\end{eqnarray}    
The bare Green function is defined as, 
\begin{eqnarray}
\hat{G}^{R,-1}_0 (k;\mu) = (\mu + i\delta) \hat{\gamma}_0 - 
\sum_{j=1}^3 k_{j}\hat{\gamma}_{j} - m \hat{\gamma}_5
\equiv \sum_{j=0,1,\cdots,5} {\sf f}_j 
\hat{\gamma}_{j}. 
\nn 
\end{eqnarray}    
The inverse of the Green function thus determined is 
at most a linear combination of $\gamma_0$, $\gamma_5$, 
$\gamma_{15}$, $\cdots$ and $\gamma_{45}$,  
\begin{eqnarray} 
G^{R,-1}(\mu) \equiv {\sf F}_0 \hat{\gamma}_0  
+ {\sf F}_5 \hat{\gamma}_5 + \sum_{j=1}^4 {\sf F}_{j5} 
\hat{\gamma}_{j5}. \nn 
\end{eqnarray} 
Eq.~(\ref{scb1}) determines their coefficients, 
\begin{eqnarray}
&&\hspace{-2cm} {\sf F}_0 - {\sf f}_0 =  - 
\Big\{\Delta_{0,0} + \Delta_{5,5} + \sum_{j=1}^4 \Delta_{j5,j5}\Big\}\cdot  
\int d^3 k' \overline{\sf F}_0(k') \nn \\  
&&\hspace{1.5cm} - 2\Delta_{0,5} 
\int d^3 k' \overline{\sf F}_5(k') - 2 \sum_{j=1}^4 
\bigg\{ \Delta_{0,j5} \int d^3 k' \overline{\sf F}_{j5}(k') 
\bigg\}, \label{gapp1} \\   
&& \hspace{-2cm} {\sf F}_5 - {\sf f}_5 = - 
\Big\{\Delta_{0,0} + \Delta_{5,5}
- \sum_{j=1}^4 \Delta_{j5,j5}\Big\}\cdot 
\int d^3 k' \overline{\sf F}_5(k') \nn \\
&&\hspace{1.5cm} - 2\Delta_{0,5} \int d^3 k' \overline{\sf F}_0 (k') 
- 2 \sum_{j=1}^4 
\bigg\{ \Delta_{5,j5} \int d^3 k' \overline{\sf F}_{j5}(k') 
\bigg\}, \label{gapp2} \\
&& \hspace{-2cm} {\sf F}_{j5} = - 
\Big\{ \Delta_{0,0} - \Delta_{5,5} - \sum_{m=1}^4 \Delta_{m5,m5} 
\Big\} \cdot \int d^3 k' \overline{\sf F}_{j5}(k')  \nn \\
&&\hspace{-1.6cm} - 2 \Delta_{0,j5} \int d^3 k' \overline{\sf F}_{0}(k') 
- 2 \Delta_{5,j5} \int d^3 k' \overline{\sf F}_5(k') 
- 2 \sum_{m=1}^4 \bigg\{ \Delta_{m5,j5} 
\int d^3 k' \overline{\sf F}_{m5}(k') \bigg\}. \label{gapp3}   
\end{eqnarray}     
That is, the right hand sides are given by 
these coefficients themselves,   
\begin{eqnarray}
&&\hspace{-2.3cm} 
\overline{\sf F}_0(k) =  
- \frac{{\sf F}_0}{g(k)}\Big\{k^2 - {\sf F}^2_0 + {\sf F}^2_5  
+ \sum_{j=1}^4 {\sf F}^2_{j5}\Big\}, \ \ \    
\overline{\sf F}_5(k) = 
\frac{{\sf F}_5}{g(k)}\Big\{k^2 - {\sf F}^2_0 + {\sf F}^2_5 
+ \sum_{j=1}^4 {\sf F}^2_{j5}\Big\}, \nn \\   
&& \hspace{-1.52cm} 
\overline{\sf F}_{j5}(k) =  
- \frac{{\sf F}_{j5}}{g(k)}\Big\{k^2 + {\sf F}^2_0 - {\sf F}^2_5  
-  \sum_{m=1}^4 {\sf F}^2_{m5}\Big\} - 
\big(1-\delta_{j4}\big) \frac{2k_j}{g(k)} 
\Big\{\sum_{m=1}^3 {\sf F}_{m5} k_m\Big\}, \nn  
\end{eqnarray} 
with $g(k)$ being defined as,  
\begin{eqnarray}
&&\hspace{-1.5cm}  
g(k) \equiv \bigg\{- {\sf F}^2_0 + {\sf F}^2_5 
+ \sum_{j=1}^4 {\sf F}^2_{j5} - k^2 
\bigg\}^2 + 4 k^2 \Big\{{\sf F}^2_5 - {\sf F}^2_0\Big\} + 
4 \bigg\{\sum_{m=1}^3 {\sf F}_{m5} k_m\bigg\}^2. \nn
\end{eqnarray}
Note that all integrals in eqs.~(\ref{gapp1},\ref{gapp2})  
depend on the ultraviolet cutoff; $\int d^3 k \equiv \int_{|k|<\Lambda} d^3 k$.  
       
We will solve these coupled integral equations, assuming 
that {\it the spatial inversion symmetry is recovered 
after the quenched average}. Namely, we assume   
\begin{eqnarray}
\hspace{1.6cm} 
\hat{\gamma}_5 \cdot \hat{G}^{R(A)}(k;\mu)\cdot \hat{\gamma}_5 
= \hat{G}^{R(A)}(-k;\mu), \label{inv} 
\end{eqnarray}
because only $\gamma_5$ anticommutes with both 
$\gamma_{1,\cdots,4}$ and $\gamma_{15,\cdots,45}$. 
This requires 
$\Delta_{0,j5}$, $\Delta_{5,j5}$ and ${\sf F}_{j5}$ 
$(j=1,\cdots,4)$ to be zero, so that the  
equation becomes, 
\begin{eqnarray}
&&\hspace{-2.5cm}
\left[\begin{array}{c} 
F_0 \\ 
F_5 \\
\end{array}\right]
+ G\left[\begin{array}{cc} 
\Delta_{s} + \Delta_{a} & - B \\ 
B & - \Delta_{s} + \Delta_{a} \\
\end{array}\right] 
\left[\begin{array}{c}
F_0 \\
F_5 \\
\end{array}\right] = 
\left[\begin{array}{c}
\mu \\
 - m \\
\end{array}\right], \   
G = 2 \int_{0}^{1}
\frac{k^2 dk} {F^2_0 - F^2_5 - k^2}, \label{gap3}   
\end{eqnarray} 
with $\Delta_{s} \equiv 2\pi \Lambda (\Delta_{0,0} + \Delta_{5,5})$,  
$\Delta_{a} \equiv \sum_{j=1}^{5} 2\pi \Lambda \Delta_{j5,j5}$, and  
$B=4\pi \Lambda\Delta_{0,5}$. In eq.~(\ref{gap3}), the momenta and 
energies are rescaled by the ultraviolet cut-off $\Lambda$, 
\begin{eqnarray}
\hspace{-2.3cm}
\Lambda \rightarrow 1, \ 
 k \rightarrow k\Lambda^{-1}, \ 
{\sf F}_{0,5} \rightarrow F_{0,5} 
\equiv {\sf F}_{0,5}\Lambda^{-1}, \  
(\mu,m)_{\rm old}
 \rightarrow (\mu,m)_{\rm new}
 \equiv (\mu,m)_{\rm old}\Lambda^{-1}. \label{rescale}
\end{eqnarray} 
The first part of eq.~(\ref{gap3}) can 
be `diagonalized' by the following canonical 
transformation,   
\begin{eqnarray}
\hspace{-1.7cm} 
\left[ \begin{array}{c}
{\cal F}_0 \\
{\cal F}_5 \\ 
\end{array} 
\right] = \left[ \begin{array}{cc} 
{\rm ch} \theta & -{\rm sh} \theta \\
- {\rm sh} \theta & {\rm ch} \theta \\  
\end{array} 
\right]  \left[ \begin{array}{c}
F_0 \\
F_5 \\ 
\end{array} 
\right], \ \  \left[ \begin{array}{c}
\psi_1 \\
\psi_2 \\ 
\end{array} \right] = \left[ 
\begin{array}{cc} 
{\rm ch} \theta & -{\rm sh} \theta \\
- {\rm sh} \theta & {\rm ch} \theta \\  
\end{array} 
\right]  \left[ \begin{array}{c}
\mu \\
- m \\ 
\end{array} \right], \label{cano1}  
\end{eqnarray} 
where the angle $\theta$ is chosen as 
\begin{eqnarray}
\big( {\rm ch} \theta , {\rm sh} \theta \big) \equiv   
\frac{\big(\Delta_s + \sqrt{\Delta^2_s-B^2},B\big)}
{\sqrt{2}\{\Delta^2_s-B^2\}^{\frac{1}{4}}
\{\sqrt{\Delta^2_s - B^2} + \Delta_s\}^{\frac{1}{2}}}. 
\label{bog}  
\end{eqnarray} 
Under this transformation, we obtain, 
\begin{eqnarray}
\hspace{-1.4cm}
\left\{ \begin{array}{l}
(1+\eta_{+} G) {\cal F}_0 = \psi_{1} \\
(1-\eta_{-} G) {\cal F}_5 = \psi_{2}, \\
\end{array}\right. \ \ G = 2 \int_{0}^{1} 
\frac{k^2 dk} {{\cal F}^2_0 - {\cal F}^2_5 - k^2} \equiv  
2\int_{0}^{1} \frac{k^2 dk}{(a+ib)^2 - k^2}, \label{gap3a}     
\end{eqnarray}
with $\eta_{\pm} = \sqrt{\Delta^2_s-B^2}\pm  \Delta_a$ 
and  
$(a+ib)^2\equiv {\cal F}^2_0-{\cal F}^2_5$.  
The real/imaginary part of `$G$' is an even/odd function 
of both $a$ and $b$, 
\begin{eqnarray}
&&\hspace{-1.6cm} {\rm Re}G = -2 - \frac{a}{2}{\rm log}
\Bigg[\frac{(1-a)^2+b^2}{(1+a)^2+b^2}\Bigg] 
+ b \bigg( {\rm arctan}\Big[\frac{1-a}{b}\Big] + 
{\rm arctan}\Big[\frac{1+a}{b}\Big]\bigg), \label{reg} \\
&&\hspace{-1.6cm} {\rm Im}G = - \frac{b}{2}{\rm log}
\Bigg[\frac{(1-a)^2+b^2}{(1+a)^2+b^2}\Bigg] 
- a \bigg({\rm arctan}\Big[\frac{1-a}{b}\Big] + 
{\rm arctan}\Big[\frac{1+a}{b}\Big]\bigg). \label{img} 
\end{eqnarray}  

Eqs.~(\ref{gap3a}-\ref{img}) can be solved in a quite 
similar fashion as we did in Ref~\cite{prb-sm} (compare 
them with eqs.~(38-41) in this reference). As a 
result, the final self-consistent Born phase diagram 
obtained in this section is basically same as 
that obtained there. The only difference is 
that the phase boundaries which separate 
two gapped phases and compressible (diffusive) 
phase are deformed in the $\mu$-$m$ parameter space,  
according to the canonical 
transformation eq.~(\ref{cano1}): they become 
symmetric with respect to $\psi_1$ and $\psi_2$, 
instead of $\mu$ and $m$ 
(compre Fig.~\ref{pd2} with Figs.~9,10 
in the reference). Those who are well-informed 
of Ref.~\cite{prb-sm} therefore might as well 
skip the remaining part of 
this section and start from sec.~5. To make this article 
self-contained, we will review the arguments in the context 
of eqs.~(\ref{gap3a}-\ref{img}).

\subsection{$\psi_{1}=\psi_{2}=0$ case}  
Consider the simplest case first, $\mu=m=0$. Since 
$\eta_{+}+\eta_{-}\equiv 2\sqrt{\Delta^2_s-B^2}\neq 0$
\footnote{The convergence of the gaussian integral in 
eq.~(\ref{que}) requires $\Delta_s > B$.}, the 
possible solutions of eqs.~(\ref{gap3a}-\ref{img}) 
are clearly three-folded: 
\begin{eqnarray}
\hspace{1.5cm}
\left\{ \begin{array}{l}
{\rm (i)}: \ {\cal F}_0 = {\cal F}_5 = 0, \\
{\rm (ii)}:\  1+\eta_{+} G = 0 \cap {\cal F}_5 = 0, \\
{\rm (iii)}:\  1-\eta_{-} G = 0 \cap {\cal F}_0 = 0. \\ 
\end{array} \right. \label{gap5}
\end{eqnarray}
The type-(iii) 
solution can realize when $\eta_{-}<-\frac{1}{2}$. 
In reality, however, such a parameter region is very limited and 
the solution itself is also physically unlikely. 
Thus, we will investigate only the first two types. 

The type-(i) solution 
is a diffusionless solution, where the massless zero-energy 
state is free from any disorders.
The type-(ii) solution describes the diffusive massless  
state: the zero energy state at the critical point 
acquires a finite life-time due to the non-magnetic disorders. 
The gap equations have an intrinsic critical 
disorder strength $\eta_{+,c}$, below which  
only the type-(i) solution is allowed. When the disorder 
strength exceeds this critical value, $\eta_{+}>\eta_{+,c}$, 
this type-(i) solution becomes unphysical and  
type-(ii) solution becomes a unique physical 
solution. These two solutions are continuously 
connected at $\eta_{+}=\eta_{+,c}$. 

To see this situation, begin with the 
type-(ii) solution: 
\begin{eqnarray}
\hspace{1cm} 1 + \eta_{+} {\rm Re}G = 0 \ \cap \ {\rm Im} G = 0  
\ \cap \  {\cal F}_5 = 0. 
\end{eqnarray}
${\rm Im}G=0$ requires either (a) $a=0$ or 
(b) $b=0 \cap |a|>1$. Now that `$a$' and `$b$' as well as `$F_0$' and 
`$F_5$' are rescaled by the ultraviolet 
cutoff $\Lambda$ as in eq.~(\ref{rescale}), 
they should be much smaller than unit. 
This leads to $a=0$. With this, $1+\eta_{+} {\rm Re}G=0$ 
determines `$b$' as a function of the 
disorder strength $\eta_{+}$:
\begin{eqnarray}
\hspace{1.5cm} b \cdot {\rm arctan}[b^{-1}] = 1 - \frac{1}{2\eta_{+}}. \label{gap5} 
\end{eqnarray}
This equation dictates that `$b$' can be finite 
only when $\eta_{+}$ is greater than some critical 
value. That is $\eta_{+,c}=\frac{1}{2}$. Above this, the 
type-(ii) solution becomes possible, $({\cal F}_0,{\cal F}_5) = (ib,0)$.   
When the disorder strength falls below this critical value, 
the solution reduces continuously to the trivial one, 
$({\cal F}_0,{\cal F}_5)=(0,0)$.   
In the followings, we will summarize the behaviours of these 
two solutions in the presence of finite $\psi_{1}$ and $\psi_{2}$.   
  
\subsection{$\psi_{1}\ne 0$ and $\psi_{2}= 0$ case}
Introduce small $\psi_1$ into the type-(i) solution first.  
Employing ${\cal F}_{5}=0$, we have only to solve the first 
line of eq.~(\ref{gap3a}) in favor for $a+ib={\cal F}_0$.  
Since its real part, ${\cal F}^{\prime}_{0}$, is an odd function of 
$\psi_{1}$ and its imaginary part 
${\cal F}^{\prime\prime}_{0}$ is even in $\psi_{1}$, 
`$a$' is the first order in small $\psi_{1}$ while 
`$b$' is the second order,  
$a = {\cal O}(\psi_{1})$ and $b= {\cal O}(\psi^2_1)$. 
With this, eq.~(\ref{gap3a}) can be solved iteratively in 
small $\psi_{1}$: 
\begin{eqnarray} 
a = \frac{\psi_{1}}{1-2\eta_{+}} + {\cal O}(\psi^3_1), \ \ \   
b = \frac{\pi\eta_{+}}{(1-2\eta_{+})^3}\psi^2_{1} 
+ {\cal O}(\psi^4_1). \label{asym1} 
\end{eqnarray} 
This indicates that, for $\eta_{+}>\eta_{+,c}$, 
the sign of the renormalized chemical potential `$a$' and 
that of the bare chemical potential `$\psi_1$' becomes opposite 
with each other. This is, however, physically unlikely. When 
$\eta_{+}$ exceeds this critical strength, 
the type-(ii) solution instead of type-(i) 
becomes a physical solution. Indeed, having a finite `$b$' already at 
the zero-th order in $\psi_{1}$, the type-(ii) solution behaves 
as follows, 
\begin{eqnarray}
a = \frac{\psi_{1}}{2\eta_{+}-1} + {\cal O}(\psi^3_1, b^2 \psi_{1}), 
\ \ b = \frac{1}{\pi}\frac{2\eta_{+}-1}{\eta_{+}} + {\cal O}(\psi^2_1,b^2),  
\label{asym2} 
\end{eqnarray}
in small $\psi_1$ region. For $\eta_{+}>\eta_{+,c}$, 
`$a$' has the same sign as that of the bare one.  
  
\subsection{$\psi_{1}= 0$ and $\psi_{2}\ne 0$ case}   
In the presence of finite $\psi_{2}$, 
the solutions of eqs.~(\ref{gap3a}-\ref{img}) are two-folded, 
depending on the relative ratio between the disorder strength and  
the topological mass $\psi_2$.  When the topological mass 
is less than a certain critical value, say $\psi_{2,c}$, the system 
is in the diffusive (compressible) phase, which is basically  
equivalent to the type-(ii) solution at $\psi_2=0$,    
\begin{eqnarray}
\hspace{0.5cm} 
{\cal F}_0 = i 
\sqrt{\tau^{-2}-\Big(\frac{\eta_{+}\psi_{2}}{\eta_{+}+\eta_{-}}\Big)^2}, 
\ \  {\cal F}_5 = \frac{\eta_{+}\psi_{2}}{\eta_{+} + \eta_{-}}. \label{sol0} 
\end{eqnarray}  
$\tau^{-1}$ and $\psi_{2,c}$ are defined by $\eta_{\pm}$:
\begin{eqnarray}
\hspace{0.5cm} 
\psi_{2,c}(\eta_{+},\eta_{-}) = \frac{\eta_{+} + \eta_{-}}{\eta_{+}\tau}, \ \  
\tau^{-1}{\rm arctan}[\tau] = 1 - \frac{1}{2\eta_{+}}. \label{bound} 
\end{eqnarray}  
When $\psi_{2}$ exceeds this critical value, the system 
enters one of the two incompressible phases, which is adiabatically 
connected to the gapped phases in the clean limit,    
\begin{eqnarray}
\hspace{1.5cm} 
{\cal F}_0 = 0, \ \ \ {\cal F}_5 = \overline{\psi}_2. \label{sol1}
\end{eqnarray}    
The renormalized mass terms $\overline{\psi}_2$ 
is given by 
\begin{eqnarray}
\hspace{0.7cm} 
\overline{\psi}_2 
\big\{ 1 + 2\eta_{-} - 2 \eta_{-}\overline{\psi}_2 
{\rm arctan}\big[\overline{\gamma}^{-1}_2\big]\big\} = \psi_{2}. \label{sol1a} 
\end{eqnarray} 
These two solutions are connected continuously at the phase boundary,  
$\psi_{2}=\psi_{2,c}(\eta_{+},\eta_{-})$.

\subsection{$\psi_{1}\ne 0$ and $\psi_{2}\ne 0$ case} 
\begin{figure}[h]
\begin{center}
\includegraphics[scale=0.6]{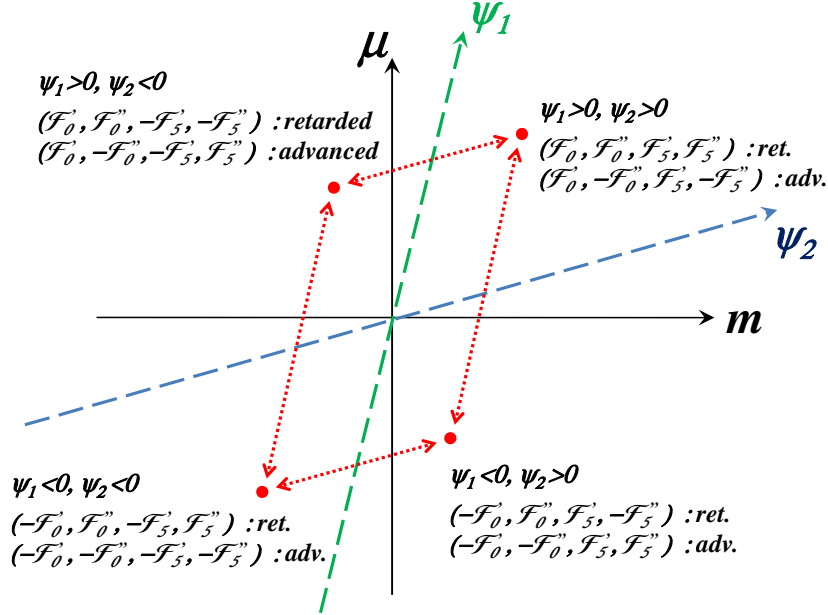}
\caption{ 
${\cal F}_{0}={\cal F}^{\prime}_0 + i{\cal F}^{\prime\prime}_0$ 
and ${\cal F}_5={\cal F}^{\prime}_5+i{\cal F}^{\prime\prime}_5$ 
as a function of bare chemical potential $\mu$ and topological mass $m$. 
${\cal F}_0$ changes its sign under 
$\psi_{1} \rightarrow - \psi_{1}$, while ${\cal F}_5$ changes 
its sign under $\psi_{2} \rightarrow - \psi_{2}$.}
\label{table1}
\end{center}
\end{figure} 

\begin{figure}[h]
\begin{center}
\includegraphics[scale=0.82]{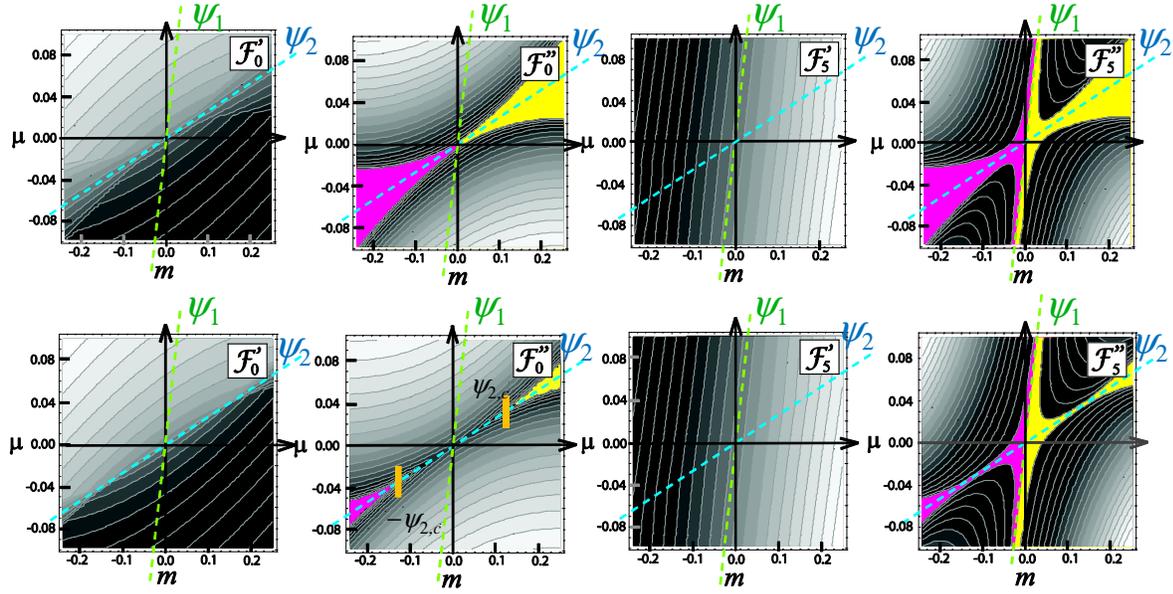} 
\caption{$({\cal F}_{0},{\cal F}_5)$ numerically 
evaluated at $\Delta_{a}/\Delta_s=B/\Delta_s=0.5$ and 
$\eta_{+}=0.45/0.57$ (upper/lower four panels).  
Their real parts and imaginary parts decrease/increase 
toward the dark/light region. 
(upper/lower four from left to right): The contour plot of  
${\cal F}^{\prime}_0$, ${\cal F}^{\prime\prime}_0$, 
${\cal F}^{\prime}_5$ and ${\cal F}^{\prime\prime}_5$ 
at $\eta_{+}=0.45/0.57$. The contour intervals are 
$0.32/0.30\times 10^{-1}$, 
$0.9/1.25 \times 10^{-2}$, $0.24/0.24 \times 10^{-1}$ and 
$0.68/0.75 \times 10^{-3}$ respectively. Both 
${\cal F}^{\prime}_0$ and ${\cal F}^{\prime\prime}_5$ 
become zero at $\psi_1=0$. ${\cal F}^{\prime\prime}_0$ 
and ${\cal F}^{\prime\prime}_5$ vanish  
within both the yellow region (topological insulator) 
and the red region (ordinary insulator).  
${\cal F}^{\prime\prime}_{0}$ in the lower panel  
remains finite even at $\psi_{1}=0$, as far as  
$-\psi_{2,c}<\psi_{2}<\psi_{2,c}$, 
where $\psi_{2,c}$ is given by eq.~(\ref{bound}).}
\label{pb}
\end{center}
\end{figure}
%
For general $\psi_{1}$ and $\psi_{2}$, we have solved 
numerically 
the following coupled equations in favor of `$a$' and `$b$': 
\begin{eqnarray}
&&\hspace{-0.7cm} 
{\cal F}_0 =  
\frac{\psi_{1} (1+ \eta_{+}{\rm Re} G)}
{(1+\eta_{+}{\rm Re} G)^2 + (\eta_{+}{\rm Im} G)^2} 
+ i\frac{\psi_{1} \eta_{+}{\rm Im} G}
{(1+\eta_{+}{\rm Re} G)^2 + (\eta_{+}{\rm Im} G)^2}, \nn \\ 
&&\hspace{-0.7cm} 
{\cal F}_5 =  
\frac{\psi_{2} (1+ \eta_{-}{\rm Re} G)}
{(1+\eta_{-}{\rm Re} G)^2 + (\eta_{-}{\rm Im} G)^2} 
+ i\frac{\psi_{2} \eta_{-}{\rm Im} G}
{(1+\eta_{-}{\rm Re} G)^2 + (\eta_{-}{\rm Im} G)^2}, \nn  
\end{eqnarray}
with 
${\cal F}^2_0-{\cal F}^2_5=(a+ib)^2$  
and ${\rm Re}G$ and ${\rm Im}G$    
defined in eqs.~(\ref{reg},\ref{img}).   
The numerical solutions thus obtained are clearly  
four-folded, i.e. $(a,b)$, $(-a,b)$, $(a,-b)$ and $(-a,-b)$.   
This leads to double degeneracy in  
${\cal F}_0$ and ${\cal F}_5$ 
\begin{eqnarray}
\hspace{1.5cm} 
({\cal F}^{\prime}_0,{\cal F}^{\prime\prime}_0, 
{\cal F}^{\prime}_5,{\cal F}^{\prime\prime}_5), \ \  
({\cal F}^{\prime}_0,-{\cal F}^{\prime\prime}_0, 
{\cal F}^{\prime}_5,-{\cal F}^{\prime\prime}_5), \nn  
\end{eqnarray} 
with ${\cal F}_0\equiv 
{\cal F}^{\prime}_0 + i {\cal F}^{\prime\prime}_0$ 
and ${\cal F}_5 \equiv 
{\cal F}^{\prime}_5 + i {\cal F}^{\prime\prime}_5$.  
Moreover, these two solutions at 
$(\psi_1,\psi_2)$ are related to those at 
the other three points, i.e. $(\psi_{1},-\psi_{2})$, $(-\psi_{1},\psi_{2})$ 
and $(-\psi_{1},-\psi_{2})$, 
\begin{eqnarray}
&&\hspace{-0.1cm} ({\cal F}^{\prime}_0,\pm {\cal F}^{\prime\prime}_0,
{\cal F}^{\prime}_5,\pm {\cal F}^{\prime\prime}_5)_{|\psi_{1},\psi_{2}} 
= ({\cal F}^{\prime}_0,\pm {\cal F}^{\prime\prime}_0,- 
{\cal F}^{\prime}_5,\mp {\cal F}^{\prime\prime}_5)_{|\psi_{1},-\psi_{2}} \nn \\ 
&&\hspace{0.1cm} = (-{\cal F}^{\prime}_0,\pm {\cal F}^{\prime\prime}_0, 
{\cal F}^{\prime}_5,\mp {\cal F}^{\prime\prime}_5)_{|-\psi_{1},\psi_{2}} 
= (-{\cal F}^{\prime}_0,\pm {\cal F}^{\prime\prime}_0,- 
{\cal F}^{\prime}_5,\pm {\cal F}^{\prime\prime}_5)_{|-\psi_{1},-\psi_{2}}.  
\label{relation}    
\end{eqnarray} 
The upper/lower sign is for the retarded/advanced Green function 
(Fig.~\ref{table1}).      

\begin{figure}[h] 
\begin{center}
\includegraphics[scale=0.75]{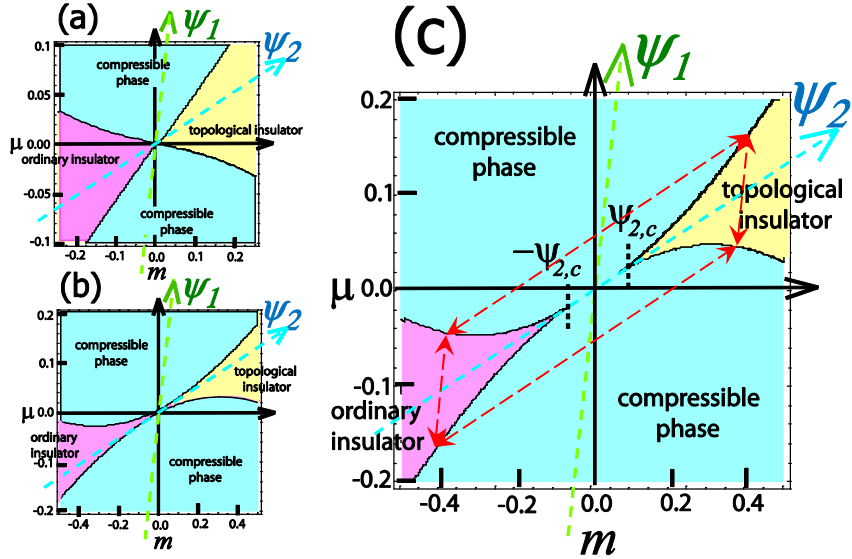}
\caption{The self-consistent Born phase diagram 
for fixed $\Delta_a/\Delta_s=B/\Delta_s=0.5$. (a/b/c) 
$\eta_{+}=0.31/0.48/0.53$, where $\eta_{+,c}=0.5$.}
\label{pd2}
\end{center}
\end{figure}

Fig.~\ref{pb} demonstrates how ${\cal F}_0$ and ${\cal F}_5$ behave 
as a function of $\mu$ and $m$, both for $\eta_{+}<\eta_{+,c}$ 
and for $\eta_{+}>\eta_{+,c}$. Fig.~\ref{pd2} is the 
corresponding phase diagrams, where both 
${\cal F}^{\prime\prime}_0$ and ${\cal F}^{\prime\prime}_5$ 
vanish in two incompressible phases (the 
topological insulator and an ordinary insulator).
For $\eta_{+}<\eta_{+,c}$, the $\mu$-$m$ phase diagram holds 
a {\it direct} phase transition point between 
these two gapped phases (Fig.~\ref{pd2}(a,b)). 
For $\eta_{+}>\eta_{+,c}$, this direct 
phase transition point becomes a {\it finite region} of 
the diffusive phase (Fig.~\ref{pd2}(c)). Especially 
at $\psi_1=0$, this diffusive region ranges from 
$\psi_2=-\psi_{2,c}$ to $\psi_2=\psi_{2,c}$, 
where $\psi_{2,c}$ is given by eq.~(\ref{bound}). 
Note also that, due to the symmetry relation given in 
eq.~(\ref{relation}), all 
the phase boundaries are symmetric with respect to the 
sign changes of $\psi_1$ and $\psi_2$ 
(see red arrows in Fig.~\ref{pd2}(c)).

 
\section{quantum conductivity correction in the presence of 
the topological-mass type disorder potential}   
We will argue the behaviour of the quantum 
conductivity correction in the presence of both chemical-potential  
type disorder and topological-mass type disorder. To 
do this, consider the following series-sum of the 
ladder type diagrams; 
\begin{eqnarray}
\hspace{-0.4cm} 
\Gamma^{\rm dif}(q,\omega) = 
\Big[\hat{\xi}^{-1} - \sum_{k}\hat{G}^{R}\big(k_{+},
\mu_{+}\big)\times \hat{G}^{A}\big(k_{-},
\mu_{-}\big)\Big]^{-1} \label{gam}
\end{eqnarray}  
where $k_{\pm}\equiv k\pm\frac{q}{2}$, 
$\mu_{\pm}\equiv \mu\pm\frac{\omega}{2}$ 
and the $16$ by $16$ tensor $\hat{\xi}$ is given by 
\footnote{Note that the product between these tensors is 
defined as $(\hat{A}_r \times \hat{A}_a) \cdot (\hat{B}_r \times \hat{B}_a) 
\equiv \hat{A}_{r}\hat{B}_r\times \hat{B}_a\hat{A}_a$, where 
$\hat{A}_r$ ($\hat{B}_r$) is a $4$ by $4$ matrix in the 
retarded line, while $\hat{A}_a$ ($\hat{B}_a$) stands for 
that in the advanced line.}; 
\begin{eqnarray}
\hspace{0.1cm}\hat{\xi} \equiv \Delta_{0,0}\hat{\gamma}_{0}\times 
\hat{\gamma}_{0} + \Delta_{0,5} \hat{\gamma}_5 \times \hat{\gamma}_0 
+ \Delta_{5,0} \hat{\gamma}_{0}\times \hat{\gamma}_5 + \Delta_{5,5} 
\hat{\gamma}_5 \times \hat{\gamma}_5. \nn 
\end{eqnarray}
Tracing out its two vertices with 
some internal degree of freedom, say $\hat{\gamma}_j$, 
we obtain the correlation function of the corresponding density  
$\psi^{\dagger}\hat{\gamma}_j \psi$, 
\begin{eqnarray}
&&\hspace{-1.4cm} 
\phi_{j}(q,\omega) \equiv - \frac{1}{2\pi i}
\sum_{\alpha,\cdots,\theta}\sum_{k,k'} 
\big[\hat{\gamma}_j\big]_{\beta\alpha} 
\hat{G}^{R}_{\alpha\epsilon}(k_{+},\mu_{+})\hat{G}^{A}
_{\zeta\beta}(k_{-},\mu_{-}) \nn \\
&&\hspace{0.9cm} 
\big\{\delta_{k,k'}
\delta_{\epsilon\delta}\delta_{\gamma\zeta}
\ + \ \Gamma^{\rm dif}_{\epsilon\theta,
\eta\zeta}(q,\omega)
\hat{G}^{R}_{\theta\delta}(k^{\prime}_{+},\mu_{+})\hat{G}^{A}
_{\gamma\eta}(k^{\prime}_{-},\mu_{-})\big\}
\big[\hat{\gamma}_j\big]_{\delta\gamma} \label{cor}
\end{eqnarray}
When this density is a conserved quantity in each ensemble, 
like $\psi^{\dagger}\hat{\gamma}_0 \psi$, the function should  
exhibit the diffusive behaviour at the infrared region, 
$q,\omega \simeq 0$. Thus, the above series-sum carries at 
least one diffusion pole.
When its advanced line is ${\cal T}$-reversed, 
it is transcribed into the series-sum of the 
`maximally-crossed' diagrams (Cooperon),  
\begin{eqnarray}
&& \hspace{-1.4cm}
(1\otimes s_y)_{\gamma\gamma^{\prime}}
\Gamma^{\rm dif}_{\alpha\delta,\beta^{\prime}\gamma^{\prime}}
(q,\omega)(1\otimes s_y)_{\beta^{\prime}\beta} 
= \sum_{i,j=0,5} \Delta_{i,j} \ [\hat{\gamma}_{i}]_{\alpha\delta} 
\times [\hat{\gamma}_{j}]_{\gamma\beta} \nn \\ 
&&\hspace{-1.6cm} + \sum_{i,\cdots,m} \sum_{k} 
\Delta_{i,j}\Delta_{l,m} \  
[\hat{\gamma}_{i}]_{\alpha\alpha_1}
\hat{G}^{R}_{\alpha_1\delta_1}(k_{+},\mu_{+}) 
[\hat{\gamma}_l]_{\delta_1\delta} \times 
[\hat{\gamma}_j]_{\gamma\gamma_1}
\hat{G}^{A}_{\gamma_1\beta_1}(-k_{-},\mu_{-}) 
[\hat{\gamma}_{m}]_{\beta_1\beta}  \nn \\ 
&&\hspace{-1.4cm} + \sum_{i,\cdots,p} \sum_{k,k^{\prime}} 
\Delta_{i,j}\Delta_{l,m} \Delta_{n,p} \ 
[\hat{\gamma}_{i}]_{\alpha\alpha_1}
\hat{G}^{R}_{\alpha_1\delta_1}(k_{+},\mu_{+})
[\hat{\gamma}_{l}]_{\delta_1\alpha_2} 
\hat{G}^{R}_{\alpha_2\delta_2}(k^{\prime}_{+},\mu_{+})
[\hat{\gamma}_{n}]_{\delta_2\delta} \nn \\ 
&& \hspace{-0.cm} 
\times [\hat{\gamma}_j]_{\gamma\gamma_1}
\hat{G}^{A}_{\gamma_1\beta_1}(-k_{-},\mu_{-}) 
[\hat{\gamma}_{m}]_{\beta_1\gamma_2}
\hat{G}^{A}_{\gamma_2\beta_2}(-k^{\prime}_{-},\mu_{-}) 
[\hat{\gamma}_{p}]_{\beta_2\beta}  + \cdots, \label{coop}   
\end{eqnarray}   
which describes the quantum interference between 
the time-reversal paired scattering process from 
$K$ to $-K+q$. Thus, any type of the diffusion pole 
appearing in eq.~(\ref{gam}) generally results in a certain 
quantum interference effect in the backward 
scattering channel.  
  
To investigate the nature of the diffusion poles in eq.~(\ref{gam}), 
employ the canonical transformation used in sec.~4 first,
\begin{eqnarray}
&& \hspace{0.8cm}
\overline{\Gamma}^{\rm dif}(q,\omega) \equiv V^{R}_{\theta} 
\times V^{A}_{\theta} \cdot \Gamma^{\rm dif}(q,\omega) \cdot 
V^{R}_{\theta} \times V^{A}_{\theta}, \nn \\ 
&&\hspace{1.8cm}
V^{R}_{\theta} = V^{A}_{\theta} = 
\cosh \frac{\theta}{2} \ 
\hat{\gamma}_0 + 
\sinh \frac{\theta}{2} \ \hat{\gamma}_5, \nn 
\end{eqnarray} 
with $\theta$ being defined as eq.~(\ref{bog}). 
Such a transformation diagonalizes $\hat{\xi}$. 
Simultaneously, it simplifies the expression of  
$\hat{G}^{R(A)}(k_{\pm},\mu_{\pm})$ in terms of 
${\cal F}_0$ and ${\cal F}_{5}$. Namely, the series-sum 
thus transformed reads     
\begin{eqnarray}
\hspace{-0.3cm}\overline{\Gamma}^{\rm dif}(q,\omega) = 
\Big[\hat{\xi}^{-1}_d - 
\sum_{k}{\cal G}^{R}\big(k_{+},\mu_{+}\big)
\times {\cal G}^{A}\big(k_{-},\mu_{-}\big)\Big]^{-1} 
\label{cano}
\end{eqnarray} 
with 
\begin{eqnarray}
&& \hspace{-0.64cm}
\hat{\xi}^{-1}_d \equiv \frac{\upsilon_{+}}{\upsilon^2_{+}-\upsilon^2_-}  
\ \hat{\gamma}_0 \times \hat{\gamma}_0 - \frac{\upsilon_-}
{\upsilon^2_{+}-\upsilon^2_{-}}\ \hat{\gamma}_5 
\times \hat{\gamma}_5, \nn \\
%
&&\hspace{-1.5cm} 
\hat{\cal G}^{R}\big(k_{+},\mu_{+}\big) 
\equiv \frac{{\cal F}_{0,+}}{{\cal F}^2_{0+} - k^2_{+} 
-{\cal F}^2_{5,+}} \hat{\gamma}_0 - 
\sum_{j=1}^3 k_{+,j} \hat{\gamma}_j - 
\frac{{\cal F}_{5,+}}{{\cal F}^2_{0,+} - k^2_{+} 
-{\cal F}^2_{5,+}} \hat{\gamma}_5, \nn \\
&&\hspace{-1.5cm} 
\hat{\cal G}^{A}\big(k_{-},\mu_{-}\big) 
\equiv  \frac{{\cal F}^{*}_{0,-}}
{({\cal F}^*_{0,-})^2 - k^2_{-} 
-({\cal F}^*_{5,-})^2} \hat{\gamma}_0 - 
\sum_{j=1}^3 k_{-,j} \hat{\gamma}_j - \frac{{\cal F}^{*}_{5,-}}
{({\cal F}^*_{0,-})^2 - k^2_{-} 
-({\cal F}^*_{5,-})^2}\hat{\gamma}_5, \nn \\ 
&&\hspace{-0.68cm} 
2{\upsilon}_{\pm}\equiv \sqrt{\Delta^2_{s}-B^2}\pm 2\pi 
(\Delta_{0,0} - \Delta_{5,5}). \nn
\end{eqnarray}
The subscript on ${\cal F}_{j}$ stands for its 
$\omega$-dependence; ${\cal F}_{j,\pm} 
\equiv {\cal F}_{j}(\mu\pm \frac{\omega}{2},m)$, 
where the $\mu$- and $m$-dependence of ${\cal F}_j$  
are determined by the preceding gap equation, 
eqs.~(\ref{gap3a}-\ref{img}). 
   
Taking $q$ to be zero, we 
get a relatively simpler 
expression for eq.~(\ref{cano}), 
\begin{eqnarray}
&& \hspace{0.4cm} 
\overline{\Gamma}^{\rm dif}(0,\omega) = 
 f^{-1}_{1} {\Gamma}_1 
+ f^{-1}_{2} {\Gamma}_2 
+ f^{-1}_{3} {\Gamma}_3 
+  f^{-1}_{4} {\Gamma}_{4}, \label{diff} \\
&& \hspace{0.4cm} 
f_{1} = a^2_1 - \delta a^2_{04} + \delta a^2_{23}, \ \ \ 
f_{2} = a^2_1 - a^2_{04} +  a^2_{23}, \nn \\ 
&&\hspace{0.7cm}
f_{3} = 9 a^2_1 - \delta a^2_{04} + \delta a^2_{23}, \ \ 
f_{4} = 9 a^2_1 - a^2_{04} + a^2_{23}, \nn     
\end{eqnarray}  
with $\delta a_{ij} \equiv a_i-a_j$, 
$a_{ij}\equiv a_i + a_j$ and 
\begin{eqnarray}
&&\hspace{-0.3cm} 
a_0 \equiv \frac{v_+}{v^2_+ - v^2_-} - 
\sum_{k} 
\frac{{\cal F}_{0,+} {\cal F}^{*}_{0,-}}
{(k^2-{\cal F}^2_{0,+}+{\cal F}^2_{5,+}) 
(k^2-({\cal F}^*_{0,-})^2+({\cal F}^*_{5,-})^2)}, \label{5a0} \\
&&\hspace{-0.3cm} 
a_1 \equiv -  
\sum_{k} 
\frac{k^2_x} {(k^2-{\cal F}^2_{0,+}+{\cal F}^2_{5,+}) 
(k^2-({\cal F}^*_{0,-})^2+({\cal F}^*_{5,-})^2)}, \label{5a1} \\ 
&&\hspace{-0.3cm} 
a_2 \equiv  \sum_{k} 
\frac{{\cal F}_{0,+} {\cal F}^{*}_{5,-}}
{(k^2-{\cal F}^2_{0,+}+{\cal F}^2_{5,+}) 
(k^2-({\cal F}^*_{0,-})^2+({\cal F}^*_{5,-})^2)}, \label{5a2} \\
&&\hspace{-0.3cm} 
a_3 \equiv \sum_{k} 
\frac{{\cal F}_{5,+} {\cal F}^{*}_{0,-}}
{(k^2-{\cal F}^2_{0,+}+{\cal F}^2_{5,+}) 
(k^2-({\cal F}^*_{0,-})^2+({\cal F}^*_{5,-})^2)}, \label{5a3} \\
&&\hspace{-0.3cm} 
a_4 \equiv - \frac{v_-}{v^2_+ - v^2_-} - 
\sum_{k} 
\frac{{\cal F}_{5,+} {\cal F}^{*}_{5,-}}
{(k^2-{\cal F}^2_{0,+}+{\cal F}^2_{5,+}) 
(k^2-({\cal F}^*_{0,-})^2+({\cal F}^*_{5,-})^2)}, \label{5a4} 
\end{eqnarray} 
Four $\Gamma_{j}$ in eq.~(\ref{diff}) 
turns out to be all regular functions (tensors) of 
$\omega$~\cite{prb-sm}. The only 
possibility is, therefore, the diffusion pole appears in 
one of the four coefficients $f_j$. 
In fact, employing the same technique invented in 
Ref.~\cite{prb-sm}, we can easily show that  
eq.~(\ref{gap3a}) requires $f_{4}$ to be zero  
at $\omega = 0$. 
The other three coefficients are generally 
truncated by some finite infrared cutoff. Each cutoff 
stands for (the inverse of) the relaxation time of a certain 
internal degree of freedom other than charge density. 

When the topological-mass-type disorder 
is absent~\cite{prb-sm}, $f_{3}$ also exhibits 
the diffusive behaviour 
at the massless point, $m=0$.  
This is because, in the presence of only the $\gamma_0$-type disorder 
potential, $\hat{\gamma}_{45}$ always commutes with   
a hamiltonian at the TQCP, so that not only 
the charge density but the corresponding parity density\footnote{
$\hat{\gamma}_{45}$ is spatially-inversion odd, while  
time-reversal even, so that we call the corresponding 
density, $\psi^{\dagger}\hat{\gamma}_{45}\psi$, 
as the parity density.} also follow a diffusion equation. 
Indeed, when substituted into eq.~(\ref{cor}), 
$f^{-1}_{3}\Gamma_{3}$ contributes to the correlation 
function of this parity density, 
while $f^{-1}_4\Gamma_4$ does to that of the charge density.  
When their advanced lines are 
time-reversed as in eq.~(\ref{coop}), both $\Gamma_3$ 
and $\Gamma_{4}$ result in the same magnitude of 
the positive weak-localization (AWL) correction 
to the charge conductivity. Thus, 
the additional $U(1)$ symmetry recovery at the TQCP 
induces the `doubling phenomena' of  
the quantum conductivity correction around this massless point. 
 

In the presence of the topological-mass-type disorder, 
however, the relaxation time of this parity 
density, given below, {\it always remains finite} in 
the {\it whole} $\mu$-$m$ parameter space, 
\begin{eqnarray}
\hspace{0.1cm}
\tau^{-1}_{\rm topo} \equiv  
\Big\{\frac{\partial f_4}{\partial \omega} 
(f_3-f_4) \Big\}_{|\omega=0} = 
\Big\{\frac{\partial f_4}{\partial \omega} 
(-a_0 a_4 + a_2 a_3) \Big\}_{|\omega=0}.  \label{topo}
\end{eqnarray}
Namely, though both $a_{2}$ and $a_3$ in the r.h.s. vanish 
at $\psi_2\equiv -\mu \sinh \theta + m \cosh \theta =0$, 
$a_4$ never vanishes even at $\psi_2=0$, due to the first 
term in eq.~(\ref{5a4}), $- \frac{v_-}{v^2_{+}-v^2_-}$.  
Indeed, once the $\gamma_5$-type impurity potential 
is introduced, the parity density is generally a 
non-conserved quantity. As a result, unlike in Ref.~\cite{prb-sm}, 
only the charge diffusion pole results in the positive 
weak-localization correction, while that of the parity 
diffusion mode remains always `gapped' in the 
entire $\mu$-$m$ parameter space.  
 

\section{summary} 
From the bulk-edge correspondence, the topological 
quantum critical point (TQCP) which intervenes the three-dimensional 
topological insulator and an ordinary 
insulator is expected to be stable against any non-magnetic 
disorders, provided that each surface state supported in the 
topological insulator phase is stable against the same 
perturbations. To understand the stability of this 3-d TQCP, 
we first employ 
the `Berry phase' argument and show that 
any backward scattering process which conserves  
the parity density degree of freedom, 
$\psi^{\dagger}\gamma_{45}\psi$, is  
always set off by its ${\cal T}$-reversal 
counter-process. This observation upholds 
more directly our previous surmise in 
Ref~\cite{prb-sm}, where 
two of the authors conjectured that, when the system transits 
from the topological insulator side to the ordinary insulator 
side, the parity density always becomes a conserved quantity 
once. However, it is still an open issue {\it how} the parity 
density becomes a conserved quantity at 
these transition points (or regions) 
{\it in the presence of generic non-magnetic disorders}. 
Namely, the `Berry phase' argument {\it only} confirms 
that, as far as the parity density  
is a conserved quantity, the 3-d TQCP remains 
delocalized (or critical) even in the presence of 
sufficiently strong generic non-magnetic disorders.  

In fact, as for those backward scattering processes  
which do not conserve the parity density, the above `Berry phase' argument 
does not work at all. To understand these generic ${\cal T}$-symmetric 
situations, we further derive the self-consistent Born phase diagram in 
the presence of general non-magnetic disorders. 
We found that the derived scB phase diagram 
is basically same as that in the case with only the 
chemical-potential-type disorder. Namely, as in 
Ref.~\cite{prb-sm}, there exists a certain critical 
disorder strength, below which the two gapped phases 
(the topological insulator and an ordinary 
insulator) are always separated by the {\it direct}  
quantum phase transition point in the $\mu$-$m$ 
parameter space. When the disorder strength exceeds 
this critical value, this direct transition 
point becomes a {\it finite region} of the 
diffusive metallic phase. 

As for the quantum conductivity correction,  
the situation changes drastically. 
On increasing the topological-mass type disorder 
potential, we found that the diffusive parity 
density mode observed at the 3-d TQCP always acquires a 
{\it finite relaxation time} in the 
{\it entire} $\mu$-$m$ parameter space. This indicates that 
the `doubling phenomena' of the quantum conductivity correction 
previously observed at the TQCP~\cite{prb-sm} becomes 
suppressed in the presence of $\gamma_5$-type disorder 
potential.   


\section{Acknowledgments}
RS and RN were supported by the Institute of Physical 
and Chemical Research (RIKEN) and SM was supported 
by Grants-in-Aid for Scientific Research from 
the Ministry of Education, Culture, 
Sports, Science and Technology (MEXT) of Japan. 

\end{document}